# New Nuclear And Subnuclear Exotic Decays

**D. B. Ion[1,2], Reveica Ion-Mihai[3], M. L. Ion[3]**

[1] *Institute for Physics and Nuclear Engineering, IFIN-HH, Bucharest Romania*

[2] *TH-Division, CERN, CH-1211 Geneva 23, Switzerland*

[3] *Bucharest University, Faculty of Physics, Bucharest, Romania*

**Abstract:** *In this paper new nuclear and subnuclear exotic decays are investigated. Some theoretical problems of the pionic radioactivity, such as fission-like models, applicable to all kind of exotic nuclear and subnuclear decays are presented. The induced nuclear and subnuclear decays are discussed. Moreover, using the recent results on the spontaneous fission half lives $T_{SF}$ of the heavy nuclei with $Z \geq 100$ new predictions on the pionic yields in the region of superheavy elements are presented.*

## 1. New exotic nuclear and subnuclear radioactivities

The traditional picture of the nucleus as a collection of neutrons and protons bound together via the strong force has proven remarkable successful in understanding a rich variety of nuclear properties. However, the achievement of modern nuclear physics that not only nucleons are relevant in the study of nuclear dynamics but that pions and the baryonic resonances like $\Delta$'s and N* play an important role too. So, when the nucleus is explored at short distance scales the presence of short lived subatomic particles, such as the pion and delta, are revealed as nuclear constituents. The role of pions, deltas, quarks and gluons in the structure of nuclei is one of challenging frontier of modern nuclear physics. This modern picture of the nucleus bring us to the idea [1] to search for new exotic natural radioactivities such as: ($\pi, \mu, \Delta, N^*$, etc.)-emission during the nuclear fission in the region of heavy and superheavy nuclei. So, in 1985, we initiated the investigation of the nuclear spontaneous pion emission as a new possible nuclear radioactivity called *nuclear pionic radioactivity* (NPIR) [1-36] with possible essential contributions to the instability of heavy and superheavy nuclei [16-19, 29]. Moreover, new exotic radioactivities such as new mode of nuclear fission (*deltonic fission* and *hyperfision*), was also suggested and investigated.

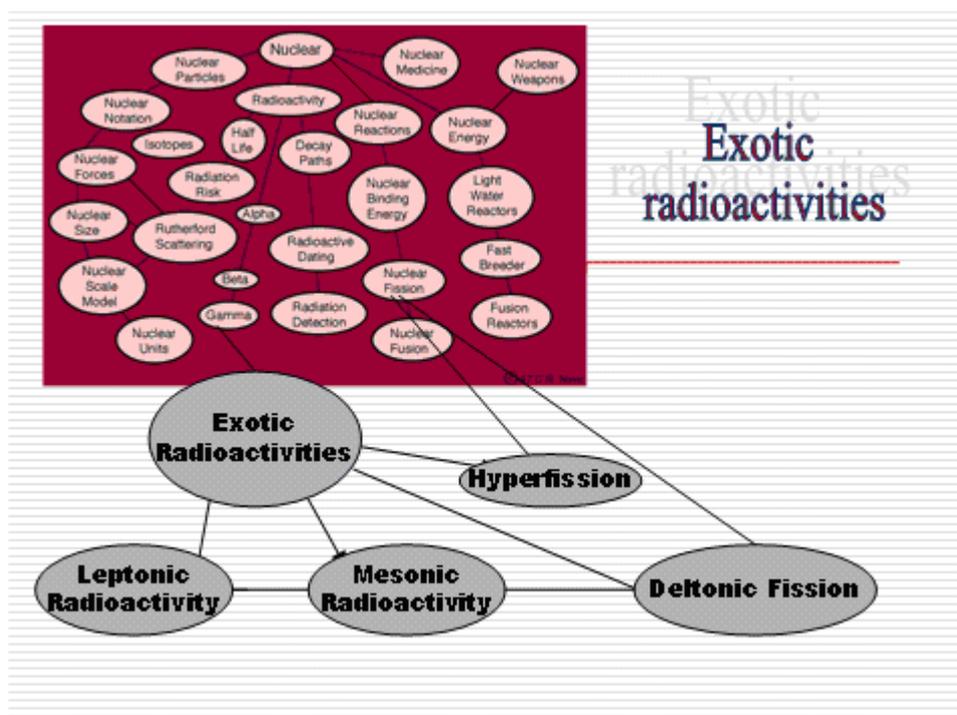

**Fig.1**: Schematic description of exotic nuclear radioactivities

The nuclear pionic radioactivity of a parent nucleus (A,Z) can be considered as an inclusive reaction of form:

$$(A,Z) \to \pi + X \tag{1}$$

where X denotes any configuration of final particles (fragments, light neutral and charged particles, etc.) which accompany emission process. The inclusive NPIR is in fact a sum of all exclusive nuclear reactions allowed by the conservation laws in which a pion can be emitted by a nucleus from its ground state. The important exclusive reactions which give the essential contribution to the inclusive NPIR (1) are the spontaneous pion emission accompanied by two body fission:

$$^A_Z X \to \pi + ^{A_1}_{Z_1} X + ^{A_2}_{Z_2} X, \text{ where } A=A_1+A_2 \text{ and } Z=Z_1+Z_2+Z_\pi \tag{2}$$

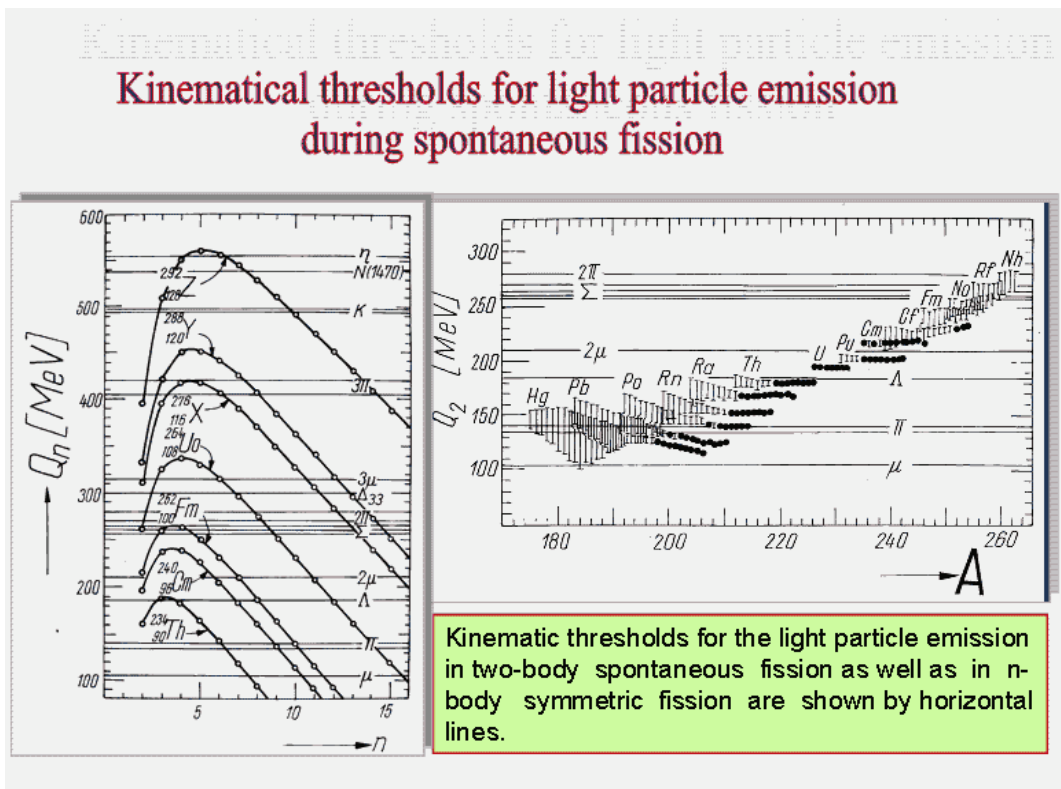

Fig. 2. $Q_n$-energies liberated in a symmetric n-body fission and $Q_2$-energies liberated in a two-body spontaneous fission. The horizontal lines correspond to the kinematical thresholds for the light particles emission during fission.

Other interesting natural radioactivities predicted at IFIN-HH-Bucharest are as follows

- ***Muonic radioactivity*** [12] is a nuclear process in which the lepton pair $(\mu^\pm, \nu_\mu)$ is emitted during two or many body fission of the parent nucleus. For the nuclear spontaneous emission of $(\mu^\pm, \nu_\mu)$ in binary fission we can write

$$^A_Z X \to \mu^\pm + \nu_\mu + ^{A_1}_{Z_1} X + ^{A_2}_{Z_2} X \, , \text{ where } A=A_1+A_2 \text{ and } Z=Z_1+Z_2+Z_\mu \tag{3}$$

- ***Lambdonic radioactivity*** [13] is just a nuclear reaction of form

$$^A_Z X \to \Lambda^0 + ^{A_1}_{Z_1} X + ^{A_2}_{Z_2} X \, , \text{ where } A=A_1+A_2 \text{ and } Z=Z_1+Z_2 \tag{4}$$

- ***Hyperonic radioactivities*** *($\Sigma^-, \Sigma^0, \Sigma^+$)* are also the possible nuclear decays of form

$$^A_Z X \rightarrow \Sigma^{\pm,0} + ^{A_1}_{Z_1} X + ^{A_2}_{Z_2} X, \quad \text{where } A=A_1+A_2 \text{ and } Z=Z_1+Z_2+Z_\Sigma \qquad (5)$$

- ***Hyperfission or weak fission*** [14] is a fission process in which in one of fragment is hypernucleus:

$$^A_Z X \rightarrow ^{A_1}_{Z_1} X + ^{A_2}_{Z_2} X_Y, \quad \text{where } A=A_1+A_2 \text{ and } Z=Z_1+Z_2 \qquad (6)$$

- ***Deltonic and $N^*$ fissions*** [15] are also possible new mode of fission processes in which in one of fragment contain a delta or a $N^*$ resonance:

$$^A_Z X \rightarrow ^{A_1}_{Z_1} X + ^{A_2}_{Z_2} X_\Delta, \quad \text{where } A=A_1+A_2 \text{ and } Z=Z_1+Z_2 \qquad (7)$$

Hence, the *nuclear pionic radioactivity* (1)-(2), $(\mu^\pm, \Lambda^0, \Sigma^{\pm,0}, \text{etc.})$-radioactivities (3)-(5), as well as hyperfissions and $(N^*, \Delta)$-new modes of fission, are extremely complex coherent reactions in which we are dealing with a spontaneous particle emission accompanied by a rearrangement of the parent nucleus in two or many final nuclei. Charged pions as well as neutral pions can be emitted during two body or many body fission of the parent nucleus. New spontaneous as well as induced exotic radioactivities can be introduced as described in Fig.3.

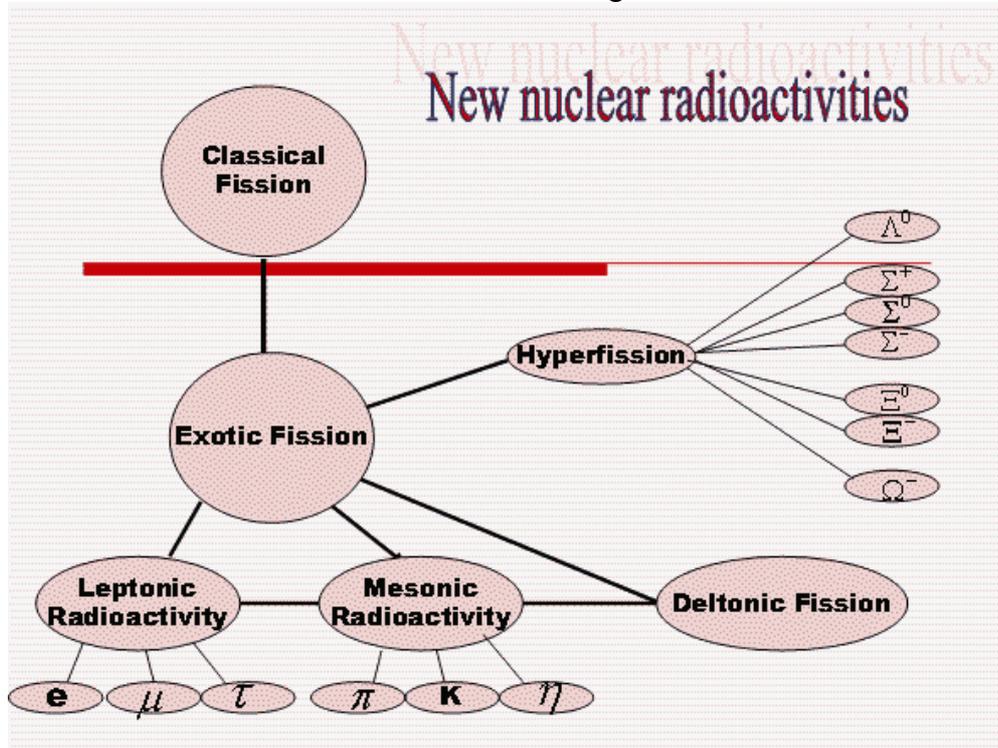

Fig. 3: New exotic nuclear radioactivities

## 2. The fission-like model for exotic nuclear and subnuclear radioactivities

A fission-like model [5, 20, 30, 55] for the pionic radioactivity was regarded as a first stage in the development of an approximate theory of this new phenomenon that takes into account the essential degree of freedom of the system: $\pi - fissility$, $\pi - fission\ barrier\ height$, etc.
Therefore, let us consider

$$E_C^{\pi F}(0) = E_C^0 - \alpha m_\pi \qquad (8)$$

$$E_S^{\pi F}(0) = E_S^0 - (1-\alpha) m_\pi \qquad (9)$$

where $E_C^0$ and $E_S^0$ are the usual Coulomb energy and surface energy of the parent nucleus given by

$$E_C^0 = \gamma Z^2 / A^{1/3} \quad \text{and} \quad E_S^0 = \beta A^{2/3} \qquad (10)$$

with $\beta = 17.80$ MeV and $\gamma = 0.71$ MeV. $\alpha$ is a parameter defined so that $\alpha m_\pi$ and $(1-\alpha)m_\pi$ are the Colombian and nuclear contributions to the pion mass (for $\alpha = 1$, the entire pion mass is obtained only from Coulomb energy of the parent nucleus). So, by analogy with binary fission was introduced the pionic fissility $X_{\pi F}$ which is given by

$$X^{(\alpha)}_{\pi F} = \frac{E_C^{\pi F}(0)}{2E_S^{\pi F}(0)} = \frac{E_C^0 - \alpha m_\pi}{E_S^0 - (1-\alpha)m_\pi}, \quad 0 \leq \alpha \leq 1 \qquad (11)$$

or

$$\left(\frac{Z^2}{A}\right)_{\pi F} = \frac{Z^2}{A} - \frac{m_\pi}{\gamma A^{2/3}} \frac{\alpha - (1-\alpha)E_C^0/E_S^0}{1 - (1-\alpha)m_\pi/E_S^0} \qquad (12)$$

In the particular case $\alpha = 1$ we have

$$X_{\pi F} = X_{SF} - \frac{m_\pi}{2E_S^0} \quad \text{or} \quad \left(\frac{Z^2}{A}\right)_{\pi F} = \frac{Z^2}{A} - \frac{m_\pi}{\gamma A^{2/3}} \qquad (13)$$

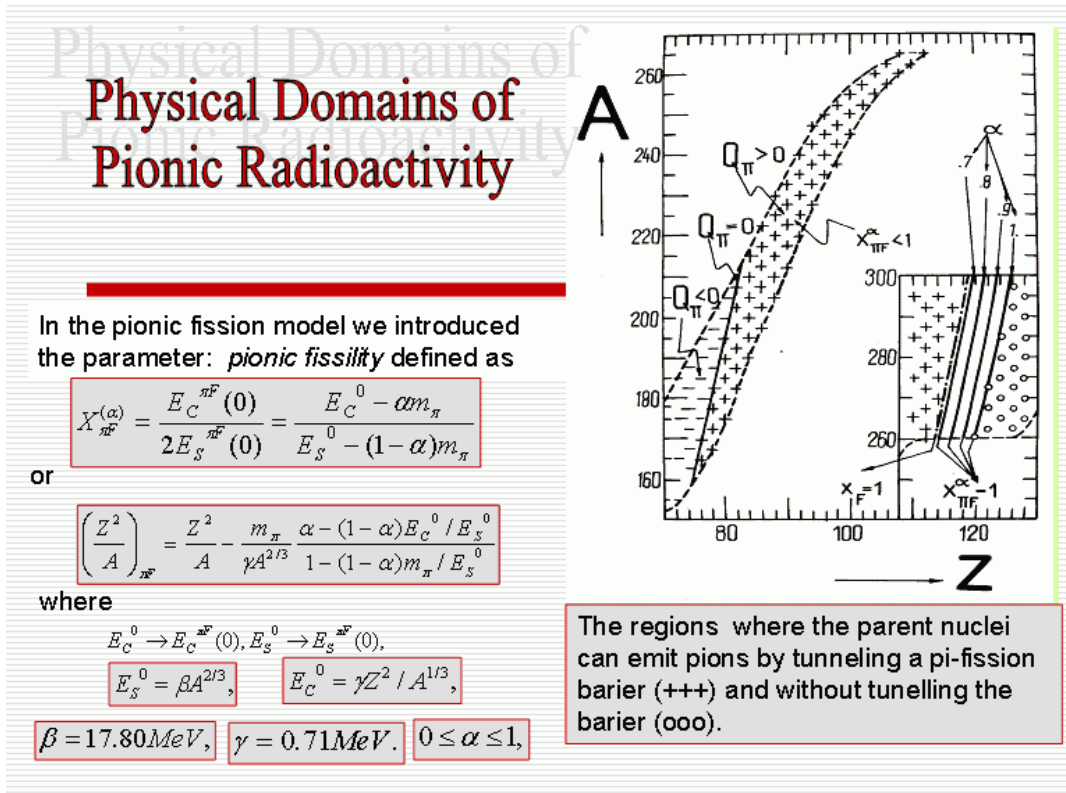

Fig.4: Physical regions for the pionic radioactivity

In Fig. 4 we presented the regions from the plane (A, Z) in which some parent nuclei are able to emit spontaneously pions during the nuclear spontaneous fission.
Therefore, according to Fig. 4, we have the following important regions:
- SHE (super heavy elements)-region, indicated by white circles, where $X_{\pi F} > 1$ and $Q_\pi > 0$, all the nuclei are able to emit spontaneously pions during the SF since no pion fission barrier exists.
- HE (heavy elements)-zone marked by signs plus (+++), corresponding to $X_{\pi F} < 1$ and $Q_\pi > 0$, where all the nuclei can emit spontaneously pion only by quantum tunneling of the pionic fission barrier.
- E-region, indicated by signs minus (----) where the spontaneous pion emission is energetically interdicted since $Q_\pi < 0$.

## Pionic fission barrier

Pionic fission barier is defined by the substitutions

$$E_C^0 \to E_C^{\pi F}(0), \quad E_S^0 \to E_S^{\pi F}(0)$$

$$E_C^{\pi F}(0) = E_C^0 - \alpha m_\pi \qquad E_S^{\pi F}(0) = E_S^0 - (1-\alpha)m_\pi$$

Then, we obtain

$$\Delta E^{\pi F} = \Delta E_C^{\pi F} + \Delta E_S^{\pi F}$$
$$= \frac{\varepsilon^2}{5}[2E_S^{\pi F}(0) - E_C^{\pi F}(0)]$$
$$= \frac{2\varepsilon^2 E_S^{\pi F}(0)}{5}[1 - X^{(\alpha)}{}_{\pi F}]$$

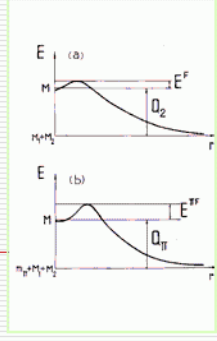

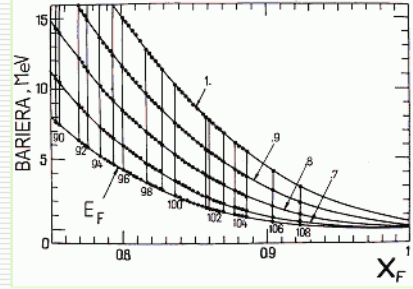

Definition of pionic-fission barier

Fig. 5: Barrier height for pionic fission

## Nuclear configuration at "sadle" point

$$R(\theta) = \frac{R_0}{\lambda}[1 + \varepsilon_2 P_2(\cos\theta) + \varepsilon_4 P_4(\cos\theta) + \varepsilon_6 P_6(\cos) + ...]$$

$$\varepsilon_2 = 2.33(1-X_{\pi F}) - 1.23(1-X_{\pi F})^2 + 9.50(1-X_{\pi F})^3 - 8.05(1-X_{\pi F})^4 +$$

$$\varepsilon_4 = 1.98(1-X_{\pi F}) - 1.70(1-X_{\pi F})^2 + 17.74(1-X_{\pi F})^4 + ..$$

$$\varepsilon_6 = -0.95(1-X_{\pi F}) + ...$$

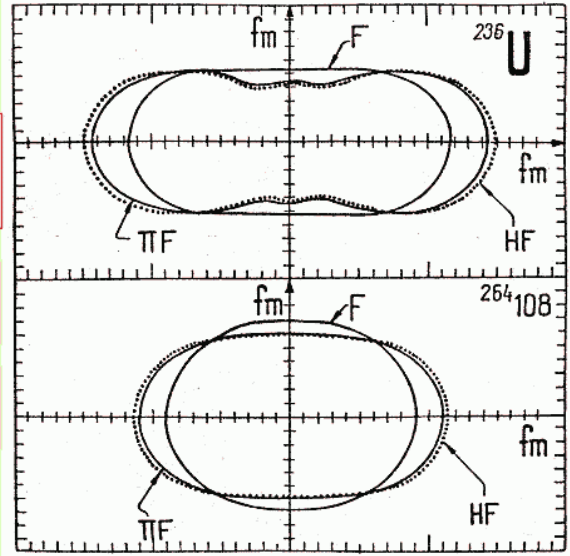

Fig. 6: The nuclear configuration at the saddle point for spontaneous fission (F), spontaneous hyperfission (HF) and spontaneous pionic fission ($\pi F$)

We note that the above definitions are valid only with the constraints

$$E_C^{\pi F}(0) + E_S^{\pi F}(0) + m_\pi = E_C^0 + E_S^0 \qquad (14)$$

They are applicable to all kind of exotic nuclear and subnuclear decays with the substitution:

$$m_\pi \Rightarrow \Delta m_x \qquad (15)$$

where $\Delta m_x$ is the energy necessary to create the x-particle on mass shell.

The *dynamical thresholds* for the pionic fission are obtained just as in fission case by using the substitution:

$$X_{SF} \to X_{\pi F}, \quad E_C^0 \to E_C^{\pi F}(0), E_S^0 \to E_S^{\pi F}(0) \tag{16}$$

Hence, by analogy with binary fission the *barrier height* for the pionic fission in a liquid drop model can be written as:

$$E^{\pi F}(LD) = E_S(0)[0.73(1-X_{\pi F})^3 - 0.33(1-X_{\pi F})^4 + 1.92(1-X_{\pi F})^5 - 0.21(1-X_{\pi F})^6] \tag{17}$$

In Fig. 5 and 6 we present the values of barrier height $E^{\pi F}(LD)$ as well as the nuclear configuration *at the saddle point compared with those from fission (F) or hypefission (HF)*.

In Fig. 6 $R_0$ is the spherical radius and $\lambda$ is a scale factor just as in binary spontaneous fission. Of course, the true barrier height for the pionic fission

$$E^{\pi F} = E^{\pi F}(LD) - \Delta E_{shell}^{\pi F}(s.p.) - \Delta E_{shell}(g.s.) \tag{18}$$

where $\Delta E_{shell}^{\pi F}(s.p.)$ and $\Delta E_{shell}(g.s.)$ are correction due to shell effect at saddle point and ground state, respectively.

Detailed predictions for the pionic radioactivity are obtained [4-7, 29, 33] by assuming a scaling law for any kind of fission accompanied by the light particle emission.

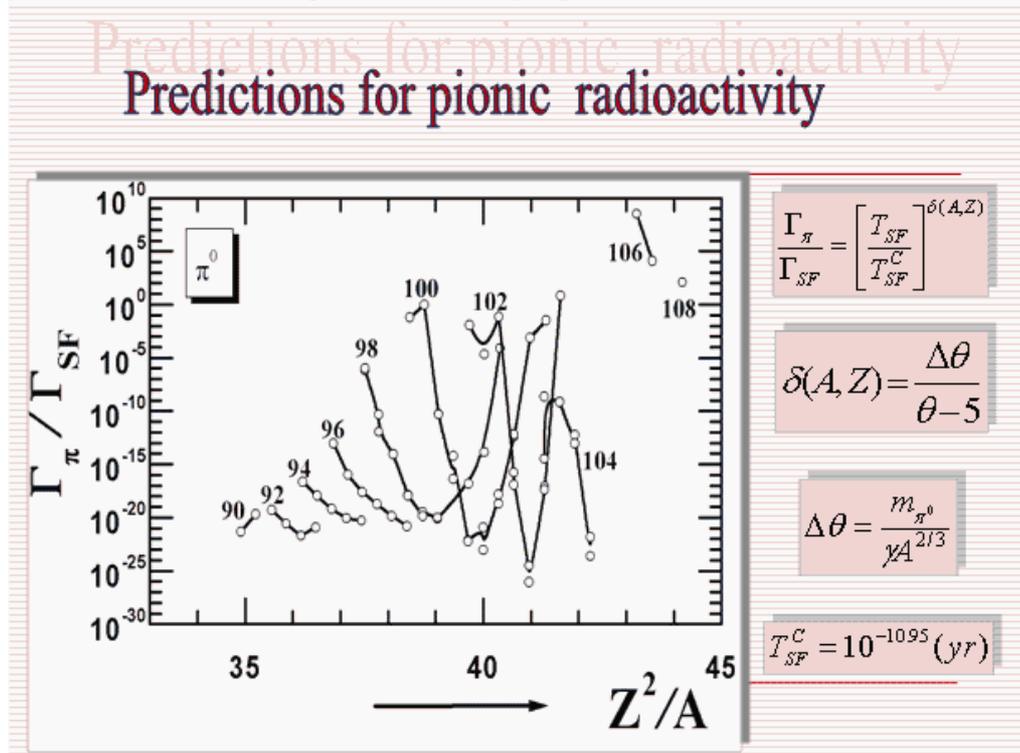

Fig. 7: Predictions for $\pi^0$ – spontaneous emission during fission of heavy and superheavy nuclei.

## 3. Supergiant pionic radioactive halos

The radioactive halos were first reported between 1880 and 1890 and their origin was a mystery until the discovery of radioactivity and its power of coloration. In 1907 Joly and Mugge, independently, suggested that the central inclusion of a halos was radioactive and that alpha-emission from it produced spherically concentric shells of coloration. Therefore, aside from their interest as attractive mineralogical oddities, the halos are of great interest for the nuclear physics because they are an integral record of radioactive decay in minerals. This integral record is detailed enough to allow estimation of the energy involved in the decay process and to identify the decaying nuclides through genetic connections. This latter possibility is particularly exciting because there exist certain classes of halos, such as the dwarf halos, the X-halos, the giant halos and the supergiant halos, which cannot be identified with the ring structure of the known alpha-emitters. Hence, barring the possibility of a non radioactive origin, these new variants of halos can be interpreted as evidences for hitherto undiscovered alpha-radionuclide, as well as, signals for the existence of new types of radioactivities. In the spirit of these ideas based on the pionic radioactivity hypothesis in the paper [30] the pionic radiohalos (PIRH), as the integral record in time of the pionic nuclear radioactivity of the heavy nuclides with Z>80 from the inclusions from ancient minerals, are proposed. Then, it is shown that the essential characteristic features of the observed supergiant halos (SGH) observed by Grady and Walker [51] and Laemmlein [52], are reproduced with a surprising high accuracy by those of the pionic radiohalos.

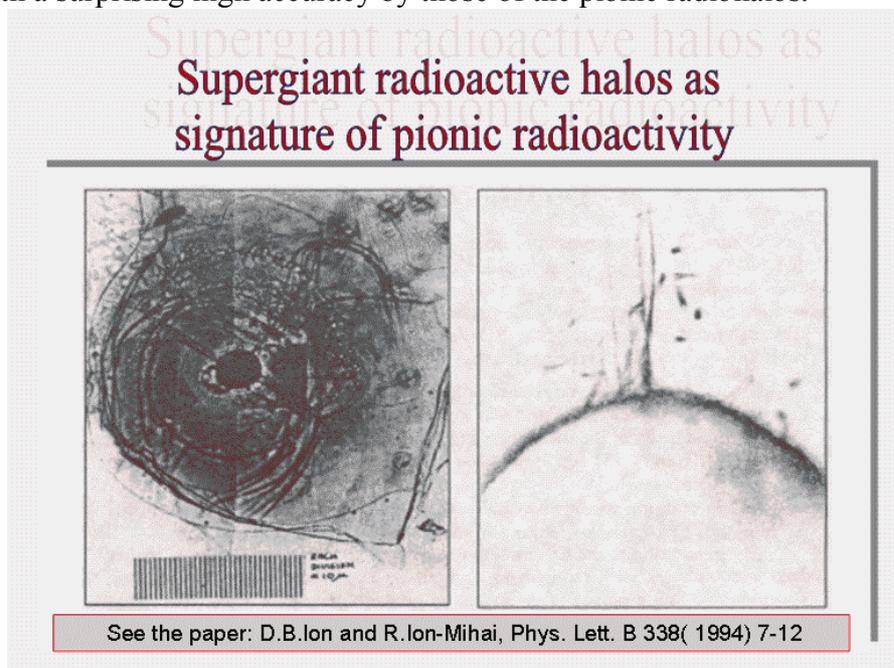

Fig. 8. A photomosaic of a part of one supergiant halos (SGH) F-12 discovered by Grady and Walker [51].

## 4. Experimental evidences

Recently Khryachkov et al. [45] presented a new spectrometer for investigation of ternary nuclear fission. The measured characteristics of this spectrometer allow for its successful use in studies of ternary fission with the emission of $\alpha$-particles, tritons, and protons as well as in the search for exotic nuclear fission accompanied by the emission of charged mesons $(\pi^{\pm}, \mu^{\pm})$. This new spectrometer was tested with a reaction of spontaneous $^{252}$Cf ternary fission. The measurements were carried out continuously 1.5 months. A spectrum of the scintilator signals obtained in coincidence with fragments is shown in Fig.9 (see Fig. 9 in Ref. [45]).

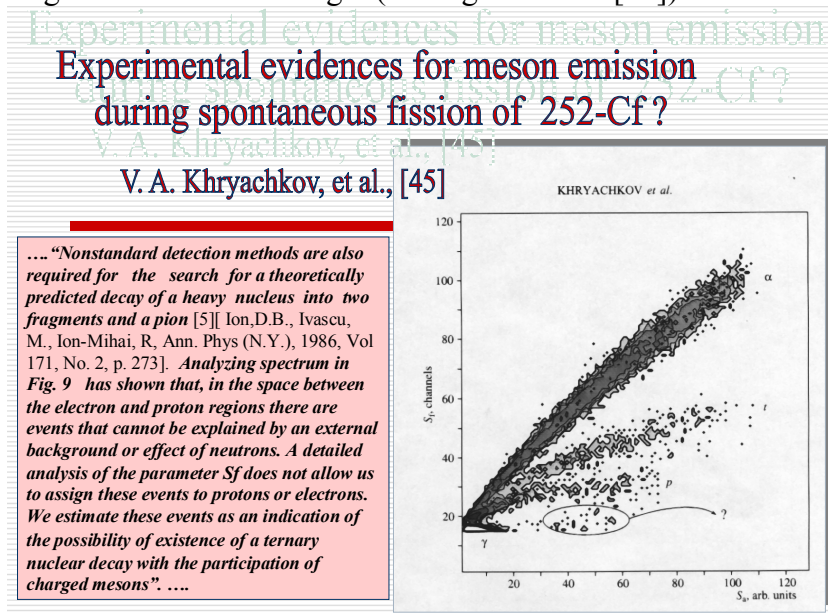

**Fig. 9:** The experimental results (see Fig. 9 in Ref. [45]) on the light charged particle emission during the spontaneous fission of $^{252}$Cf**.**

We note that the results from Fig. 9 are in a qualitative agreement with the results presented by us in Fig.3 of the paper [22].

# 5. Induced exotic radioactivities.

The investigation of the induced exotic radioactivities (see Fig. 10) are also of fundamental importance for the nuclear physics.

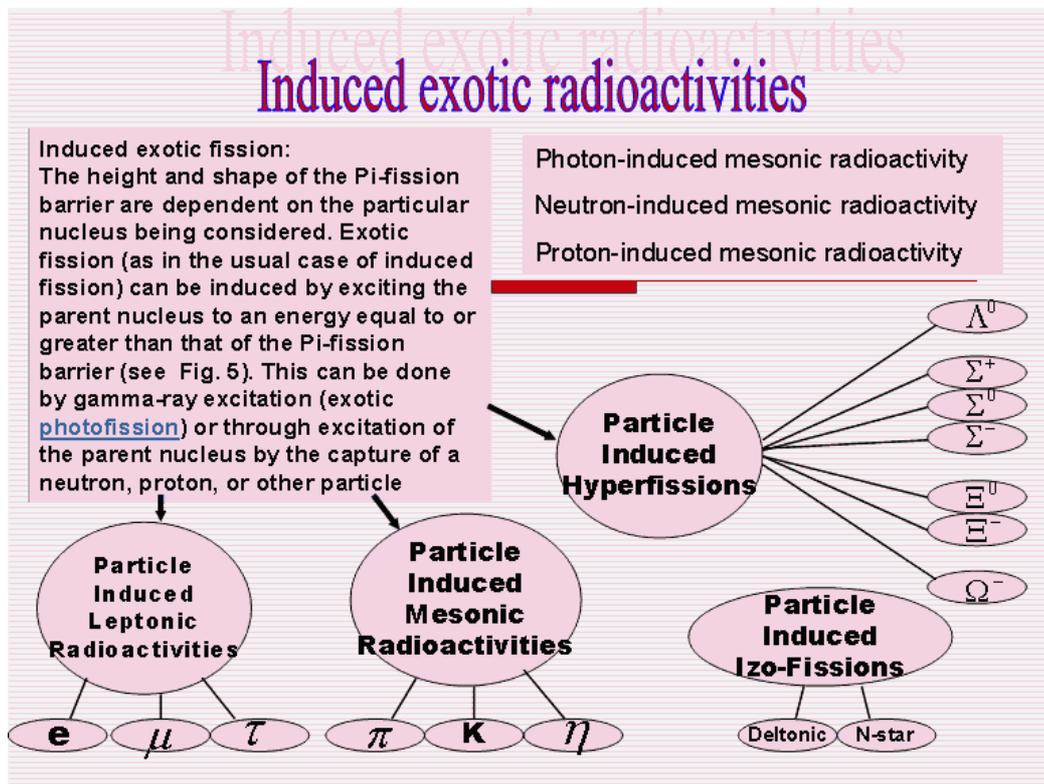

Fig. 10. Schematic description of possible new exotic induced radioactivities.

Induced nuclear fission accompanied by pion emission was investigated in few papers [22], [53]-[54]. Obviously, the probability of the induced $\pi - fission$ channel can be many orders of magnitude larger than that in spontaneous $\pi - fission$ channel.

In particular, the induced pionic radioactivities can also be of great importance not only for nuclear fundamental physics but also for their possible applications. By anticipating, the discovery of the induced pionic fission with rapid neutrons, in the future decades, we expect that our results on the $\pi - fission$ barrier height will contribute in essential and decisive way to the discovery of new domain of the nuclear physics, namely, the domain of *the induced pionic radioactivities.* For the planed investigations on proton induced pionic fission see the "Site-ul" of the Institute of Nuclear Research of Russian Academy of Science (IYaI RAN) at the address:
http://www.inr.ru/rus/mmfideljap.html.

# 6. Summary and Conclusions.

In this paper new nuclear and subnuclear exotic decays are investigated. Some theoretical problems of the pionic radioactivity, such as fission-like models, applicable to all kind of exotic nuclear and subnuclear decays are presented.

The experimental and theoretical results obtained on the NPIR in more than 20 years can be summarized as follows:
1. A fission-like model (see Ref.[5]) for the pionic radioactivity was regarded as a first stage in the development of an approximate theory of this new phenomenon that takes into account the essential degree of freedom of the system: $\pi - fissility$ , $\pi - fission$ *barrier height, nuclear configuration at* $\pi F - sadle$ *point,* etc. Detailed predictions for the pionic radioactivity (see Ref. [4-7,29,33] are obtained by assuming a scaling law for any kind of fission accompanied by the light particle emission. Some new predictions on the pionic yelds in the region of SHE-nuclei are presented in Fig. 7.

2. A statistical model [8] as well as a detailed balance model [9] for pionic radioactivity is presented.
3. A new interpretation of the experimental bimodal fission in terms of the unitarity diagrams is obtained in Refs. [31, 34]. So, the presence of the symmetric mode in the fragment mass-distribution at transfermium nuclei can be interpreted as experimental signature of the pionic radioactivity. Then, it is expected that this new degree of nuclear instability can becomes dominant at SHE-nuclides [16, 17].
4. The nuclear pionic radioactivity (NPIR) was experimentally investigated by many authors [23]-[28], [36]-[47]. A short review of the experimental limits obtained on the spontaneous NPIR yields is presented in Table 3 of Ref. [33]. The best experimental limit for $\pi^0$-yields has been reported for $^{252}$Cf by Bellini et al. [40]. They reached an upper limit of $3.10^{-13}$, a value close to the theoretical prediction [5].
5. The unusual background, observed by Wild et al.[50] in ($\Delta$E-E)-energy region below that characteristic for long range alpha emission from $^{257}$Fm was interpreted by Ion, Bucurescu and Ion-Mihai [28] as being produced by negative pions emitted spontaneously by $^{257}$Fm. Then, they inferred value of the pionic yield is: $\Gamma_\pi/\Gamma_{SF} < (1.2 \pm 0.2).10^{-3}$ $\pi^-$/fission. In a similar way, Janko and Povinec [43] obtained the yield: $\Gamma_\pi/\Gamma_{SF} < (7 \pm 6).10^{-5}$ $\pi^+$/fission.
6. The pionic $\pi^\pm$ – supergiant radiohalos are introduced by us in [30] (see also [32,35]). Then, it is shown [30] that the supergiant radiohalos (SGH), discovered by Grady and Walker [51] and Laemmlein [51] can be interpreted as being the $\pi^-$ – $radihalos$ and $\pi^+$ – $radihalos$, respectively. Hence, these supergiant radiohalos can be considered as experimental evidences of the integral record in time of the natural pionic radioactivity from radioactive inclusions in ancient minerals.
7. The pionic radioactivity as dominant decay mode of superheavy elements (SHE) was investigated in Refs [16-19, 29]. It is shown that the so called island of stability around the magic nucleus $^{298}$[114] is destroyed due to the dominant pionic radioactivity channel. New island of stability against the pionic decay mode are presented: for even-even nuclei the predicted stability island is around the nuclei $^{304}$[120] $T_{\pi F} \approx 10^{2.5} yr$, while for A-odd the predicted island is around the nucleus $^{303}$[119] with $T_{\pi F} \approx 10^6 yr$.

Finally, we note that more dedicated experiments using nuclei with theoretically predicted high pionic yields, are desired since the discovery of the nuclear pionic radioactivity is of fundamental importance in nuclear science not only for the clarification of high instability of SHE- nuclei but also for new nuclear applications.